\documentstyle[12pt,epsfig]{article}
\newfont{\cmu}{cmu10 scaled\magstep1}
\newcommand{\be}{\begin{equation}}
\newcommand{\ee}{\end{equation}}
\newcommand{\bea}{\begin{eqnarray}}
\newcommand{\eea}{\end{eqnarray}}
\newcommand{\PD}{{\partial}}

\begin{document}
\begin{flushright}
UFTP--477/1998\\
YRHI-98-11\\
nucl-th/9806010\\
\end{flushright}
\vspace*{2cm}
\begin{center}
{\Large\bf Entropy Production in Collisions of Relativistic Heavy Ions -- 
a signal for Quark-Gluon Plasma phase transition?}
\\[.5cm]
{\bf M.\ Reiter${}^a$, A.\ Dumitru${}^b$, J.\ Brachmann${}^a$,}\\
{\bf J.A.\ Maruhn${}^a$, H.\ St\"ocker${}^a$, W.\ Greiner${}^a$}
\\[0.2cm]
{\small ${}^a$Institut f\"ur Theoretische Physik der J.W.\ Goethe-Universit\"at}\\
{\small Robert-Mayer Str. 10, Postfach 11 19 32}\\
{\small D-60054 Frankfurt a.M., Germany}
\\[0.4cm]
{\small ${}^b$Physics Department, Yale University}\\
{\small P.O. Box 208124, New Haven, Connecticut 06520-8124, USA}
\\[1cm]
{June 2, 1998}
\end{center}
\begin{abstract}
Entropy production in the compression stage of heavy
ion collisions is discussed within three distinct macroscopic models (i.e.\
generalized RHTA, geometrical overlap model and three-fluid hydrodynamics).
We find that within these models $\sim$80\% or more of the experimentally
observed final-state entropy is created in the early stage. It is thus
likely followed by a nearly isentropic expansion.
We employ an equation of state with a first-order phase transition. For
low net baryon density, the entropy density exhibits a jump at the phase
boundary. However, the excitation function of the
specific entropy per net baryon, $S/A$, does not reflect this jump.
This is due to the fact that for final states (of the
compression) in the mixed phase, the baryon
density $\rho_B$ increases with $\sqrt{s}$, but not the temperature $T$.
Calculations within the three-fluid model show that a large fraction of
the entropy is produced by nuclear
shockwaves in the projectile and target.
With increasing beam energy, this fraction of $S/A$ decreases.
At $\sqrt{s}=20$~AGeV it is on the order of
the entropy of the newly produced particles around midrapidity.
Hadron ratios are calculated for the entropy
values produced initially at beam energies from $2$ to $200$ AGeV.
\end{abstract}
\vfill
\begin{flushleft}
{\cmu Work supported by BMBF, DFG, GSI and the German 
Academic Exchange Service (DAAD).}
\end{flushleft}
\newpage

\section{Introduction}

A large amount of entropy can be produced in the initial off-equilibrium
stage of energetic nuclear
collisions \cite{entropy}.
The subsequent expansion is, on the other hand, often assumed to be 
isentropic. The linear increase of the pion multiplicity with beam
energy observed at the BEVALAC \cite{harris} is consistent with a
picture \cite{stoe,bertsch,csernai} where more than 80\% of the specific 
entropy is
produced initially. In this scenario the entropy produced during
the compression stage is closely linked to the finally observable relative
particle yields.

In the present paper, the entropy produced within the
compression
stage of highly relativistic heavy ion collisions is calculated within
three distinct macroscopic models (i.e.\
generalized RHTA, geometrical overlap model, three-fluid hydrodynamics).
Several comparisons of experimental data on relative hadron multiplicities
with calculations of chemically equilibrated hadronic gases 
indicate that the entropy per net baryon at (chemical)
freeze-out is on the order of $S/A=10\pm2.5$ at AGS and $S/A=40\pm10$ 
at SPS energies \cite{PBMAGS,CSCG,sollfrank,letessier}.
The microscopic transport model UrQMD also exhibits a 
chemical composition in the central volume of Au+Au collisions at AGS
energies which is compatible with 
an ideal hadron gas with $S/A=12$ \cite{bravina}.
Moreover, $S/A$ values in this range yield the measured
rapidity distributions and
transverse momentum spectra of produced hadrons at AGS and SPS, if an
isentropic expansion of the hot and dense matter
created during the early compression stage is assumed \cite{sollfrank2}.

The question arises, whether and how such a tremendous amount of entropy can
indeed be
created in the initial stage of the reaction, before local equilibrium is
established. If true, this would speak in favour of the picture that
the compression stage produces most of the finally observed entropy and
that it is followed by a nearly isentropic expansion.

The present study focuses on the excitation function of
$S/A$, i.e.\ its dependence on the bombarding energy. In particular,
it is interesting to study the behaviour of $S/A$ for energies where 
the matter in the central region enters the mixed phase and the pure QGP phase.
This analysis reveals whether the produced entropy exhibits an
``anomalous'' behaviour which could serve as a direct sign of the QGP phase
transition \cite{letessier,marek}. 

Also, the fraction of entropy produced both by
nuclear shockwaves in the projectile and target, on the one hand, and by
particle production (and their thermalization) around
midrapidity, on the other hand, is studied as a function of bombarding
energy within the three-fluid model.

\section{Models for the compression stage}

In this section, the equation of state and the three collective models which
will be employed
in section~\ref{res} to calculate entropy production are described.

\subsection{The Equation of State}
\label{eos}

The equation of state (EoS) of the highly excited nuclear
matter must be specified in order to determine entropy production in the
compression stage.
Here, the $\sigma - \omega$ model \cite{sigom}
is used for the interactions between the nucleons. The
parameters of the
Lagrangian are fitted such that the properties of infinite nuclear
matter in the ground state are reproduced. The pressure of an ideal
pion gas in chemical equilibrium is added to the pressure
of the interacting nucleons to account for thermal pion production.
Other particles like $\Delta$'s and $\rho$'s are not included in the
EoS, since they probably do not reach thermal and chemical equilibrium in the
early stage of the reaction.

At high temperatures or baryon-chemical potentials, a
phase transition to the quark-gluon plasma (QGP) is assumed to occur. Here,
the MIT bag model is used to calculate the EoS \cite{Bagmodel} for massless,
non-interacting
gluons and $u$, $d$ quarks. The pressure as a function of temperature $T$ and
quark-chemical potential $\mu_q$ reads \cite{Kap79}
%%%%%%%%%%%%%%%%%%%%%%%%%%%%%%%%%%%%%%%%%%%
\begin{equation}
p_{QGP} = \frac{37\pi^2}{90}T^4+T^2\mu_q^2+\frac{1}{2\pi^2}\mu_q^4-B\quad.
\label{QGPEoSp}
\end{equation}
%%%%%%%%%%%%%%%%%%%%%%%%%%%%%%%%%%%%%%%%%%
A bag parameter of $B^{1/4}=
235$ MeV is used. The two EoS are matched by Gibbs' conditions of phase
equilibrium. The resulting EoS thus exhibits -- by construction -- a
first-order phase transition
at a critical temperature (for vanishing $\rho_B$) of $T_C
\approx170$ MeV. Baryon-, entropy-, and energy-densities can be calculated
via the usual thermodynamic relations
\bea \label{QGPEoSnse}
\rho_B &=& \frac{\PD p}{\PD \mu_B}\Bigg|_T\quad,\\
s &=& \frac{\PD p}{\PD T}\Bigg|_{\mu_B}\quad,\nonumber\\
\epsilon &=& Ts-p+\mu_B \rho_B\quad,\nonumber
\eea
where the baryochemical potential $\mu_B=3 \mu_q$.
For a more detailed discussion of this EoS refer to \cite{RiShock}.

\begin{figure}[hbtp]
\vspace{-3cm}
\centerline{\hbox{\epsfig{figure=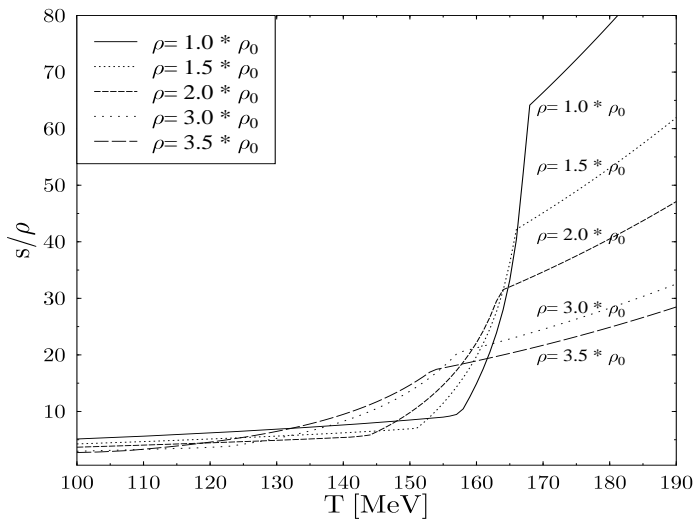,height=12cm,width=9.6cm}}}
\vspace{-2.5cm}
\caption{Entropy per net baryon as a function of temperature at various fixed
net baryon densities for the EoS used here (see text).}
\label{sT}
\vspace{.5cm}
\end{figure}

For this EoS, the entropy per net baryon $s/\rho_B$ 
exhibits a jump at the phase transition temperature, if $\rho_B$ is not
too large (Fig.~\ref{sT}). However, the
excitation function of $s/\rho_B$ does not show a corresponding jump at
a given specific collision energy, as will be discussed in section~\ref{saexcit}.

\subsection{Geometrical Overlap model}
\label{ovl}
The first model discussed is a simple schematic geometrical overlap model, where
the two Lorentz-contracted nuclei are placed on top of each other. Then
the following baryon density $\rho_B$ and energy density $\epsilon$ are
obtained:
\bea \label{overlap}
\rho_B & = & 2 \rho_0 \gamma_{CM} \\
\label{overlap2}
\epsilon & = & \sqrt{s} \rho_0 \gamma_{CM} = \frac{m_N}{2 \rho_0}
\rho_B^2\quad.
\eea
Note that the compression stage is not treated dynamically within
this model. Therefore the compression achieved is independent of the
EoS. However, the EoS determines the temperature and the entropy density of the
compressed matter.

\subsection{One-fluid hydrodynamics}
\label{Onef}
We now turn to the one-fluid hydrodynamical model, restricting
the calculations to an ideal fluid, i.e.\ viscosity and
thermoconductivity effects are neglected
\cite{csernai,danielewicz}. 

In this model, compression and entropy production occur due to 
shock discontinuities, or by a combination of shock- and isentropic
compressional
waves if the compressed matter is thermodynamically anomalous (as e.g.\
matter in the mixed phase) \cite{RiShock,Landau,Bugaev,Hofmann}. The
assumption of instantaneous thermalization leads to the maximum
energy- and baryon number-deposition in the central region that is
consistent with energy-, momentum-, and baryon number-conservation.

\begin{figure}[hbtp]
\vspace{-1cm}
\centerline{\hbox{\epsfig{figure=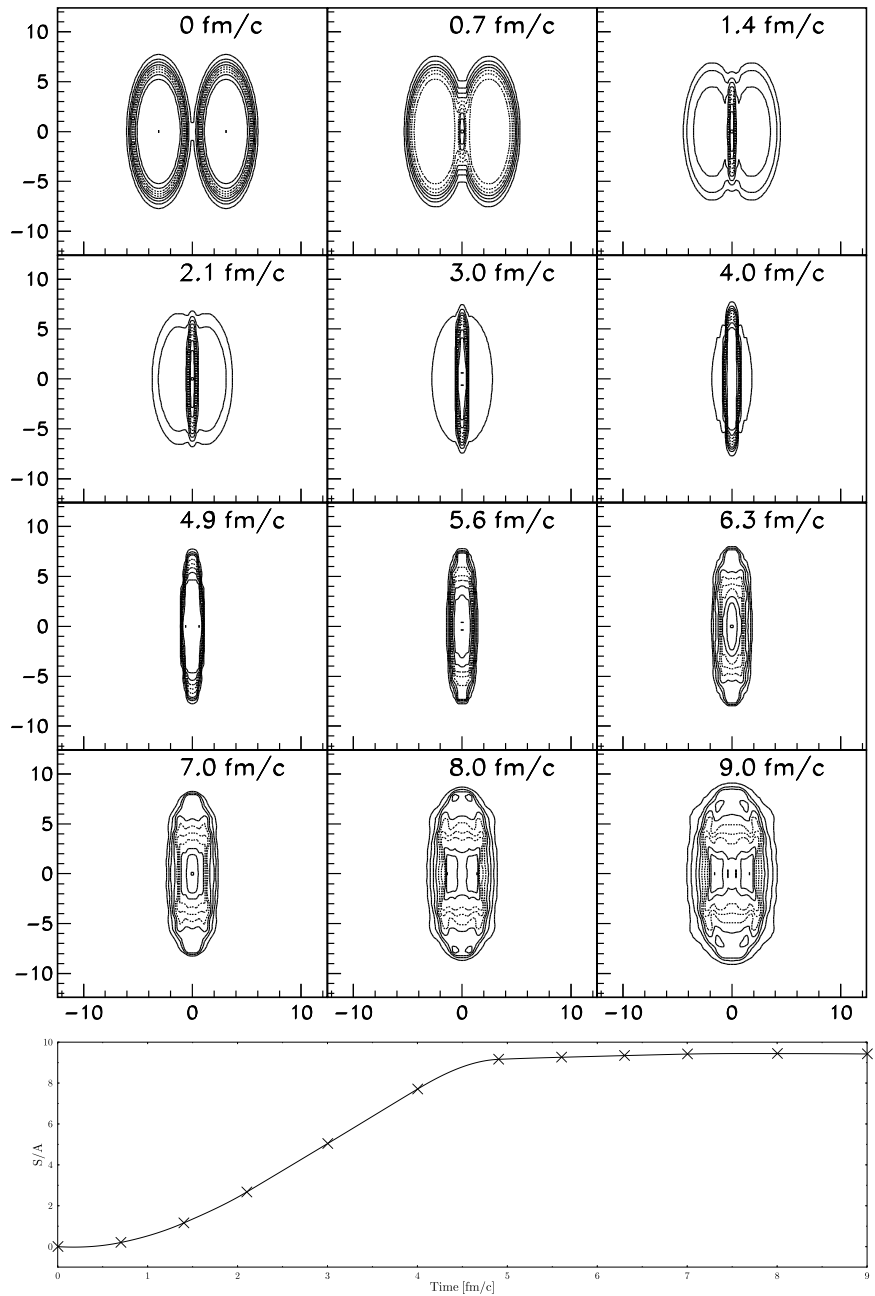,height=18cm,width=12cm}}}
\vspace{-.5cm}
\caption{Time evolution of baryon density and S/A in a 3D one-fluid
hydrodynamical calculation of Au+Au at $\protect\sqrt{s}=4.9$ AGeV.}
\label{1fsa}
\vspace{.5cm}
\end{figure}

As an example, Fig.~\ref{1fsa} shows a three-dimensional one-fluid
dynamical calculation of the collision of two Au-nuclei at
$\sqrt{s} = 4.9$ AGeV and vanishing impact parameter, $b=0$. The
upper panels of Fig.~\ref{1fsa} show reaction plane contour plots of the
baryon density $\gamma\rho_B/\rho_0$ 
in the center of mass frame. Two shockwaves can
be clearly observed as high density gradients. They develop in the center
and propagate outwards. The entropy produced by these shockwaves is 
depicted in the lower panel. The entropy per net baryon is
computed as
%%%%%%%%%%%%%%%%%%%%%%%%%%%
\be
\frac{S}{A} = \frac{\int {\rm d}^3 x \, \gamma s}{\int {\rm d}^3 x \, \gamma \rho_B}\quad ,
\end{equation}
%%%%%%%%%%%%%%%%%%%%%%%%%%%
where $s$ is the entropy density (cf.\ eq.~\ref{QGPEoSnse}).
The compression stage is finished once the shock fronts reach the back sides of the
Lorentz-contracted nuclei. Then an isentropic expansion follows (remember that we assume a
non-viscous fluid). 

The specific entropy (and also the energy density, the
pressure, etc.) of the central region (close to $x,y,z=0$) in
relativistic nucleus-nucleus collisions at $b=0$ can also be computed
directly by  solving the Rankine-Hugoniot-Taub Adiabate (RHTA) 
equation~\cite{RiShock,danielewicz,Landau,Hofmann,Taub}, thus assuming a
one-dimensional ``slab-on-slab'' collision. The RHTA determines for a
given initial state and EoS all shock wave solutions consistent with
energy and momentum conservation. Matter in the mixed phase, however, is
thermodynamically anomalous and single shock waves become mechanically
unstable. Therefore, construction of the generalized RHTA is
required~\cite{RiShock,Bugaev}. In the generalized RHTA, the unstable
shock wave solutions are replaced by a combination of shock waves and an
isentropic compression wave. For matter in the pure QGP phase, a single
shock becomes again stable.  For a given bombarding energy, the
generalized RHTA thus determines the thermodynamic properties of the
compressed matter (``final'' state). $S/A$ is independent
of $\sqrt{s}$ for all bombarding energies with
``final'' states in the phase coexistence region (cf.\
section~\ref{saexcit}), due to the fact that a simple compression wave
does not produce any additional entropy. 

\subsection{Three-fluid hydrodynamics}
\label{ThreeFluid}
The assumption of instantaneous local
thermalization in the one-fluid model prevents the expected partial
interpenetration of projectile and target.
A more realistic description, allowing for finite mean free path
effects, is obtained if two distinct fluids
for projectile and target are introduced, which interact
locally~\cite{TwoFl,pp2fluid}.
These interactions lead to energy and momentum exchange and to a smooth
deceleration of the projectile and target fluids. At AGS and SPS energies,
it is reasonable to introduce a third fluid that ``collects'' the
produced particles around midrapidity \cite{pp2fluid,BrachiNPA}. The model
assumes that the fluids are well separated in rapidity (but, of course,
not in coordinate space) during the early stage of the reaction.
Binary particle collisions increase the overlap of the fluids in
momentum space and thus drive the system into (local) mutual kinetic
equilibrium.
At this point, the various fluids merge into a single fluid
\cite{TwoFl,pp2fluid,BrachiNPA} and the subsequent expansion proceeds as in
the one-fluid
hydrodynamical model. The time scale for this kinetic equilibration of the
central
region was estimated in refs.\ \cite{BrachiNPA,Trento} to be
$\approx2R/\gamma_{CM}$.
Here the interactions between projectile and target nucleons have been regarded
as a sum over incoherent binary nucleon-nucleon scatterings (with
cross sections as in vacuum). Since we focus on the early (compressional)
stage of the reaction, the one-fluid transition is not discussed here.

The most essential difference to the one-fluid model
is that the shock fronts are smeared out considerably, i.e.\ projectile and
target
interpenetrate strongly. Consequently,
the baryons are less compressed \cite{BrachiNPA,Trento} in the
three-fluid model. Most of the energy loss of the incoming nucleons
results in particle production at midrapidity, i.e.\ a large part of the
energy loss
of the nucleons is transfered to the mid-rapidity regime (third fluid) and
only a smaller fraction of the energy is deposited
in the baryon fluids. The physical mechanism
for entropy production is very different in the
three-fluid model than in the one-fluid hydrodynamical model,
therefore the ratio of the thermal to the compression energy at midrapidity
differs from that given by the RHTA equation. 

\section{Results}
\label{res}

We now turn to the discussion of entropy production as a function of
the bombarding energy for the three different models.

\subsection{Excitation function of Specific Entropy}
\label{saexcit}

The excitation function of the entropy per (net) baryon is shown in
Fig.~\ref{sae} for the three different mo\-dels.
\begin{enumerate}
\item
The geometrical overlap model assumes that energy and baryon density increase
as $\gamma_{CM}^2$ and
$\gamma_{CM}$, respectively (cf.\ eqs.~(\ref{overlap},\ref{overlap2})). 
As a consequence,
$S/A$ increases continuously with energy.
The compressed matter enters the mixed phase at $\sqrt{s}\approx 2.4$ AGeV 
with a specific entropy of
$S/A=4.5\,$. Pure QGP matter is produced with
$S/A=12$ at $\sqrt{s}\approx 4.6$~AGeV. 
However, while sweeping through the mixed phase by increasing
the bombarding energy, $\rho_B$ and $S/A$ increase smoothly, while $T$
is roughly constant, cf.\ Fig.~\ref{Trho}.
Thus, the entropy jump seen in Fig.~\ref{sT} is not reflected in the
$S/A(\sqrt{s})$ excitation function. 
\begin{figure}[htp]
\vspace{-1cm}
\centerline{\hbox{\epsfig{figure=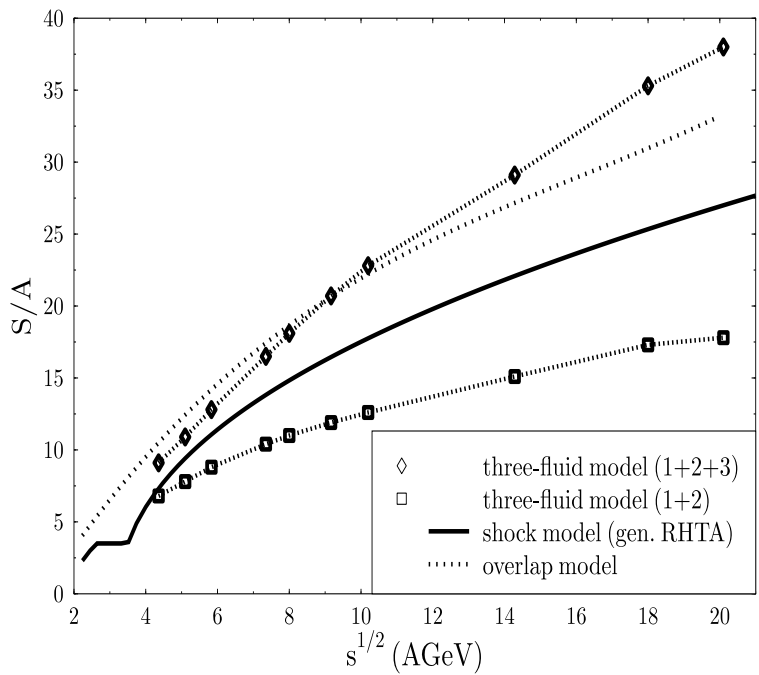,height=11cm,width=13cm}}}
\vspace{-1cm}
\caption{Excitation function of entropy per participating net baryon as
calculated within the various models. For the three-fluid model, both
the entropy of projectile- and target fluids only (1+2), as well as that of
all three fluids (1+2+3) are shown.}
\label{sae}
%\vspace{.5cm}
\end{figure}  
\item
In contrast, in the shock model (1D RHTA)
$S/A$ first increases with the beam energy and reaches a
plateau at $\sqrt{s}\approx2.7$ AGeV. At this beam energy the 
matter enters the mixed phase.  Single shock fronts cease to be the
stable solution~\cite{RiShock,Landau,Bugaev,Hofmann} in this thermodynamically
anomalous region. Rather, further compression
can only be achieved by a simple compressional wave which,
however, conserves the specific entropy. Thus, even when the
bombarding
energy is increased, $S/A$ remains constant (isentropic compression)
until the anomalous region is left again. Hence, the baryon density 
increases with energy while the
temperature {\em decreases}, as seen in
Fig.~\ref{Trho}.
\begin{figure}[hbtp]
\vspace{-3cm}
\centerline{\hbox{\epsfig{figure=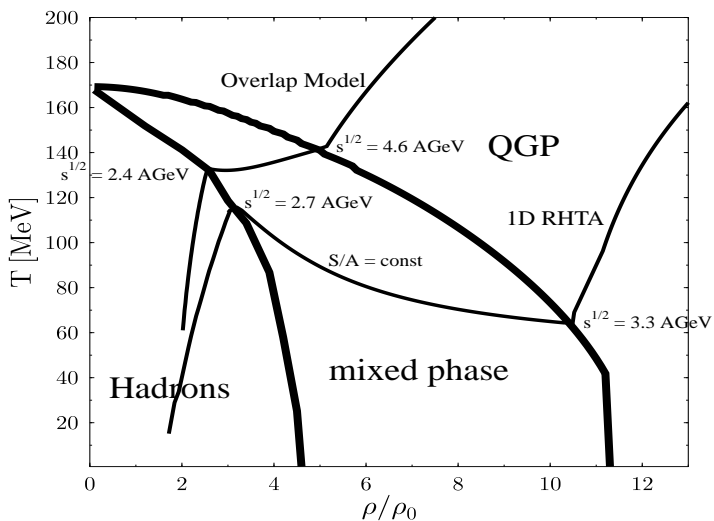,height=12cm,width=9.6cm}}}
\vspace{-2.5cm}
\caption{Temperature and (net) baryon density of the compressed matter
within the overlap model and the shock model (generalized 1D RHTA), respectively.
The phase coexistence region is also indicated.}
\label{Trho}
\vspace{.5cm}
\end{figure}  
In this model, the participant matter reaches the pure QGP phase at a
temperature of
$T=65$ MeV at $\sqrt{s}\approx3.3$ AGeV, which is significantly less than in
the overlap model (Fig.~\ref{Trho}). The
specific entropy then increases monotonically with energy. In the
AGS energy range, where a single shock is
again stable, we find $S/A\approx 10$, in agreement with the
numbers obtained from the data analysis on hadron yield ratios \cite{PBMAGS}.
However, the energy density,
baryon density, and entropy density are $\epsilon=5$ GeV/fm$^3$,
$\rho_B=2.0$ fm$^{-3}$, $s=19$ fm$^{-3}$.
These values seem unreasonably high at this energy.
These huge densities and the very early phase transition to pure QGP matter
are due to the assumption of instantaneous stopping and thermalization
\cite{TwoFl,pp2fluid,BrachiNPA}. At higher
energies, the discrepancies become even more pronounced: e.g.\ for Pb+Pb at
SPS, $\epsilon\approx50$ GeV/fm$^3$, $\rho_B\approx6$ fm$^{-3}$, and
$s\approx150$ fm$^{-3}$\/. These values
are close to the analytic expressions $\epsilon/\epsilon_0=4\gamma_{CM}^2-3$, $\rho_B/\rho_0=
4\gamma_{CM}-3/\gamma_{CM}$ ($\epsilon_0=\rho_0\cdot923$ MeV$=148$ MeV/fm$^3$,
$\rho_0=0.16$ fm$^{-3}$) derived from the RHTA with an ultrarelativistic EoS, 
$p=\epsilon/3$. Because these
densities are so huge, the specific entropy is
rather low, $S/A=25$. This is in fact significantly below the value $40\pm10$
discussed in~\cite{PBMAGS,CSCG,sollfrank,letessier,sollfrank2}.
Slightly larger $S/A$ values are found \cite{RiFri} when the number of degrees of freedom in the QGP
is increased, e.g.\ by including strange quarks.
However, at the SPS the compression phase in this model lasts only for
$\approx 1.1$ fm/$c$
(the time when the shockwaves reach the back of the nuclei). 
Within this very short time,
strange quarks may not reach thermal and chemical equilibrium. In the
subsequent expansion, strange quarks (resp.\ strange hadrons)  should
be taken into account.

\item
\begin{figure}[htp]
\vspace{-1cm}
\centerline{\hbox{\epsfig{figure=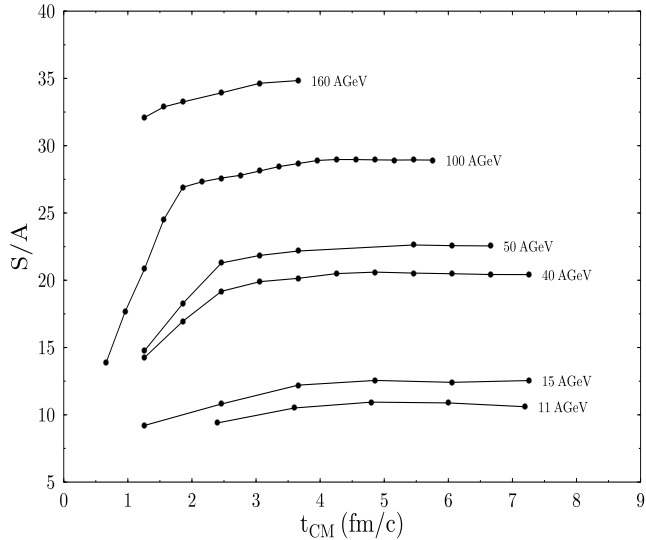,height=9cm,width=11cm}}}
\vspace{-1cm}
\caption{Time evolution of the entropy per net participant baryon 
for $Pb+Pb$ reactions at various energies, calculated within
the three-fluid model.}
\label{fig_3f}
\vspace{.5cm}
\end{figure}  
The three-fluid model predicts 
the entropy per net participating baryon as a function of
CM-time as
depicted in Fig.~\ref{fig_3f}. In contrast to our earlier work
\cite{BrachiNPA,Trento,PRCZPA}, in this calculation the EoS described in
section \ref{eos} is employed for all three fluids. The third fluid
is net baryon free since we assume that the direct baryon transfer to
midrapidity is small in the first collisions with large rapidity gap
\cite{Blobel}. 
To calculate $S/A$ within the three-fluid model a temperature cut of $T \ge
50$ MeV has been applied to determine participant baryons. This
corresponds to a mean transverse momentum of $\langle p_T \rangle \approx
270$ MeV, roughly in accordance with the $p_T$ cut employed
in~\cite{marek}. Total entropy and baryon number have then been
calculated by summing over all three fluids (resp.\ 
projectile and target fluids only for baryon number). Thus:
\be
\label{fsa3f}
\frac{S}{A} = \frac{\sum\limits_{fluids~1,2,3} \int {\rm d}^3 x \,\gamma s
\,\Theta\left(T-50~{\rm MeV}\right)}{\sum\limits_{fluids~1,2}
\int{\rm d}^3 x \,\gamma \rho_B \,\Theta\left(T-50~{\rm MeV}\right)} \quad .
\end{equation}
At early
times, $t\le2R/\gamma_{CM}$, kinetic equilibrium between the fluids is
not established. Thus, summing over the entropy and baryon currents of
the individual fluids as in eq.~(\ref{fsa3f}) seems physically more meaningful
than calculating the entropy and baryon densities from the
{\em total} energy-momentum tensor $T^{\mu\nu}=T^{\mu\nu}_1+T^{\mu\nu}_2+
T^{\mu\nu}_3$ and {\em total} baryon current $j^\mu=j^\mu_1+j^\mu_2$
{\em assuming} that they are of the form
\bea
T^{\mu\nu} &=& \left(\epsilon+p\right)u^\mu u^\nu-pg^{\mu\nu}\quad,\\
j^\mu &=& \rho_B u^\mu \quad,
\eea
as appropriate for a single (ideal) fluid \cite{Landau}.
Fig.~\ref{fig_3f} shows that
$S/A$ saturates rapidly, and is essentially time
independent for later times. This also speaks for the fact that the
numerical ``production'' of entropy is negligible in the present
calculations, which focus on the early compression phase of nuclear
collisions. Baryon number $A$ is conserved with an accuracy of $\sim 0.1
\%$ within our model, and the time independency of $S/A$ shows that this
holds also true for the entropy $S$, at least for the early stages
regarded here. 

This plateau value is plotted in
Fig.~\ref{sae} as a function of $\sqrt{s}$ (diamonds). 
If one omits the third fluid in calculating the entropy,
$S/A$ saturates similarly. The plateau values reached
in this calculation are also depicted in Fig.~\ref{sae} (squares). One observes
that with increasing bombarding energy a larger fraction of the total
entropy results from thermalization of the particles produced around mid-rapidity.
At SPS, $50$\% of the total entropy is due to shockwaves in the projectile
and target fluids.
After 
$\sim 2$ fm/$c$ (measured in the CMS) the central region is in
equilibrium \cite{Trento} and the isentropic one-fluid expansion sets in with
$S/A=38$ (for $\sqrt{s}=20$ AGeV) and $S/A=35$ (for $\sqrt{s}=18$ AGeV), respectively.
This value is $10$ units above the one-fluid result quoted above.
Consequently, particle ratios involving
antibaryons (like $\overline{N}/\pi$, $\overline{N}/N$, $\overline{\Lambda}/
\Lambda$) are expected to increase considerably, while meson-meson and baryon-baryon ratios
(like $K/\pi$, $\Lambda/N$) do not change very much
\cite{RiFri}.

Fig.~\ref{dsdy} depicts the fluid rapidity density of the entropy within
the various fluids. The fluid rapidity is defined as
\be
\label{flrap}
\tanh \eta = v_\| \quad ,
\end{equation}
where $v_\|$ is the fluid velocity component parallel to the beam axis.
One observes that the largest contribution to the entropy around $\eta =
0$ is due to production of a net baryon free QGP within the third fluid.
The entropy produced by the shockwaves in the projectile and target is
found at $\eta = \pm 1$. Note, however, that the entropy distribution in
the final state will be {\sl different} from that depicted in
Fig.~\ref{dsdy}. The high pressure gradients lead to reacceleration of
the strongly decelerated fluids, i.e.\ expansion sets in. As shown in
\cite{PRCZPA}, the resulting rapidity distribution of the pions,
including thermal smearing of the particle momenta, is Gaussian-like and
does not exhibit peaks (like ${\rm d}S/{\rm d}\eta$ at $t_{CM}=3$ fm/$c$,
cf.\ right panel of Fig.~\ref{dsdy}). In other words, at SPS energy the
three-fluid model does not predict boostinvariant initial conditions for
the expansion stage. Thus, neither the further evolution of the various
rapidity bins nor their final break-up into hadrons (cf.\
section~\ref{ratios}) proceeds independently. This is in contrast to
the dynamics at collider energies ($\sqrt{s} \ge 200$ AGeV) as discussed
in \cite{anishetty}.
\begin{figure}[thp]
\vspace{-1cm}
\centerline{\hbox{\epsfig{figure=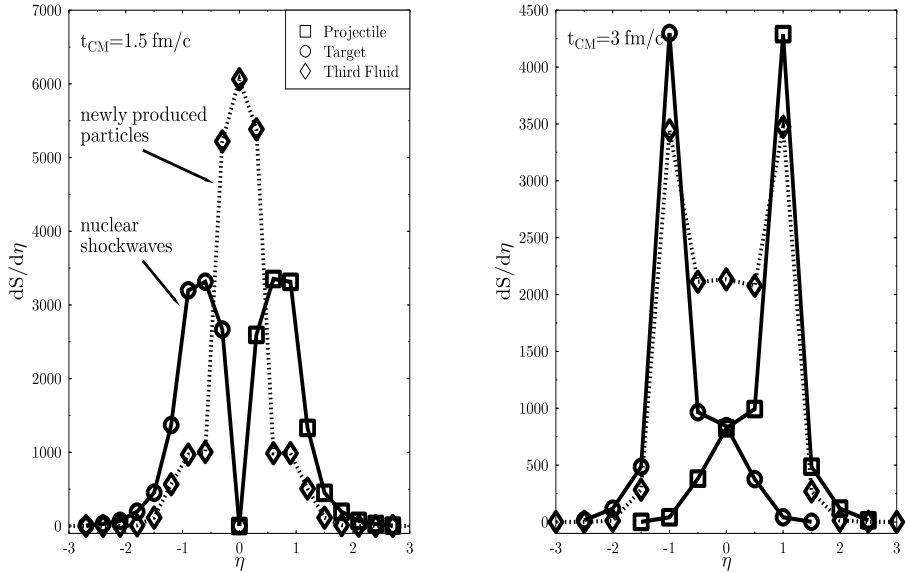,height=10cm,width=13cm}}}
\vspace{-1cm}
\caption{Fluid rapidity distribution of the entropy
in Pb+Pb at $\protect\sqrt{s}=18$ AGeV
as calculated within the three-fluid model at two different CM-times.}
\label{dsdy}
\vspace{.5cm}
\end{figure}  
\end{enumerate}

\subsection{Particle Ratios}
\label{ratios}

The initial compression stage is followed by a nearly isentropic
expansion, until the freeze-out is reached. This scenario allows to calculate
particle ratios as follows. We assume that at
(chemical) freeze-out the net baryon
density is $\rho_B^{fo}=\rho_0/2$ and the 
net strangeness of the system is zero. The third thermodynamic variable, the
specific entropy, is calculated within the three-fluid model,
cf.~section~\ref{saexcit}.
These quantities determine the chemical
composition of the fireball -- in local thermodynamical equilibrium -- unambiguously.

\begin{figure}[hbtp]
\vspace{-1cm}
\centerline{\hbox{\epsfig{figure=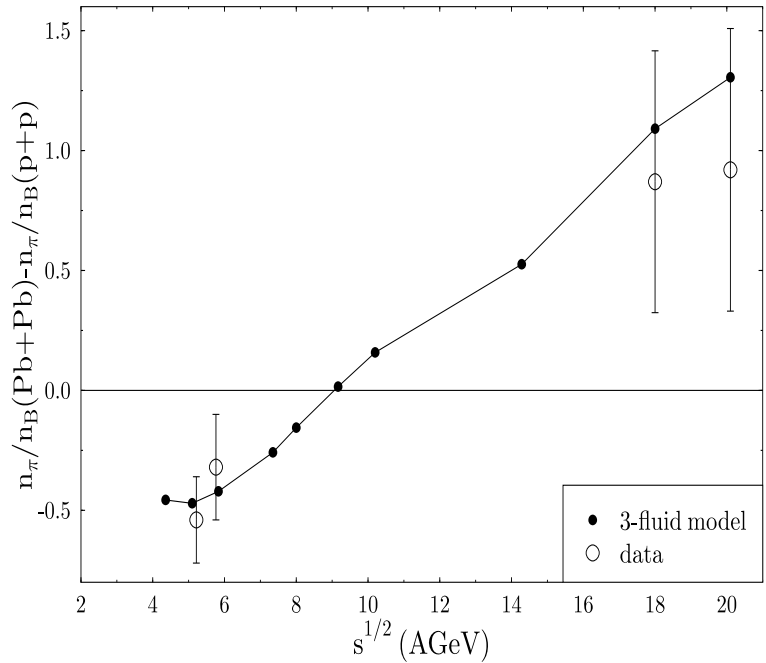,height=11cm,width=13cm}}}
\vspace{-1cm}
\caption{The number of pions per net participant baryon in Pb+Pb minus that
ratio for p+p as calculated from the S/A values obtained from the three-fluid
model. The feeding due to decays of resonances is taken into account.
Experimental data \protect\cite{marek} are also shown.}
\label{fig_pion}
\vspace{.5cm}
\end{figure}  
Investigations of hadron ratios at AGS and SPS energies
indicate that not only pions and nucleons, but also
heavier hadrons (including strange particles), and even
clusters like deuterons are close to chemical
equilibrium \cite{PBMAGS,CSCG,sollfrank,letessier} at freeze-out. 
Therefore an ideal hadron gas
model with all known hadrons and resonances up to $2$ GeV/$c^2$ mass is
employed to determine the observable
hadron ratios \cite{PBMAGS}.
Feeding from post freeze-out decays of heavy resonances
is also taken into account.

We thus implicitly assume that the (fast) compression stage is followed
by a rather long expansion stage before (chemical) freeze-out
occurs. If this is indeed the case, it is reasonable that the EoS
valid within the first $fm/c$ is different from that at the late
freeze-out stage. In principle, the chemical equilibration of
the heavier hadrons and resonances can produce additional
entropy, which is however neglected in the present analysis.

It has been pointed out that the observed number of pions per net baryon
$n_\pi/n_B$
exhibits an interesting behaviour
\cite{marek}: For collisions of nuclei in the AGS energy region
($\sqrt{s} = 5-6$ AGeV), the ratio $n_\pi/n_B$ is smaller than that in $p+p$
reactions at the same energy (per nucleon). At SPS energies ($\sqrt{s} =
18-20$ AGeV), however, the difference of the ratios is positive. 
Fig.~\ref{fig_pion} shows that this can be attributed to the much
higher specific entropy at the SPS as compared to the AGS energy. Let us
point out again that the
excitation function of this difference changes smoothly -- it 
does not exhibit a jump at some specific bombarding energy
(cf.\ sections~\ref{eos}, \ref{saexcit}) as one might expect from
Fig.~\ref{sT}. The average number
of pions in $p+p$ reactions is parametrized as a function of $\sqrt{s}$
as~\cite{antinucci}
\bea
n_{\pi^+} & = & -1.7 + 0.84 \ln s + 1.0 s^{-\frac{1}{2}}
\\ 
n_{\pi^-} & = & -2.6 + 0.87 \ln s + 2.7 s^{-\frac{1}{2}}
\\ 
n_\pi & = & \frac{3}{2} \left( n_{\pi^+} + n_{\pi^-}
\right)
\quad.\eea
At $\sqrt{s}\approx 20$ AGeV, about $50$\% of $S/A$ are due to
thermalization of the energy loss in the third fluid around mid-rapidity (at SPS,
this fluid is in
fact in the QGP phase with $\mu_q=\mu_s=0$) and about $50$\% of $S/A$
are due to nuclear shockwaves
in the projectile and target fluids (which are baryon-rich, but are
nevertheless also
in the QGP phase). Both contributions therefore are essential to understand the data. As discussed in
section~\ref{ThreeFluid}, the three fluids, however, reach kinetic
equilibrium later on and merge into a single fluid. Therefore we
consider a single fireball as the source of the frozen-out hadrons,
rather than letting each of the fluids break up independently.

\begin{figure}[hbtp]
\vspace{-1cm}
\centerline{\hbox{\epsfig{figure=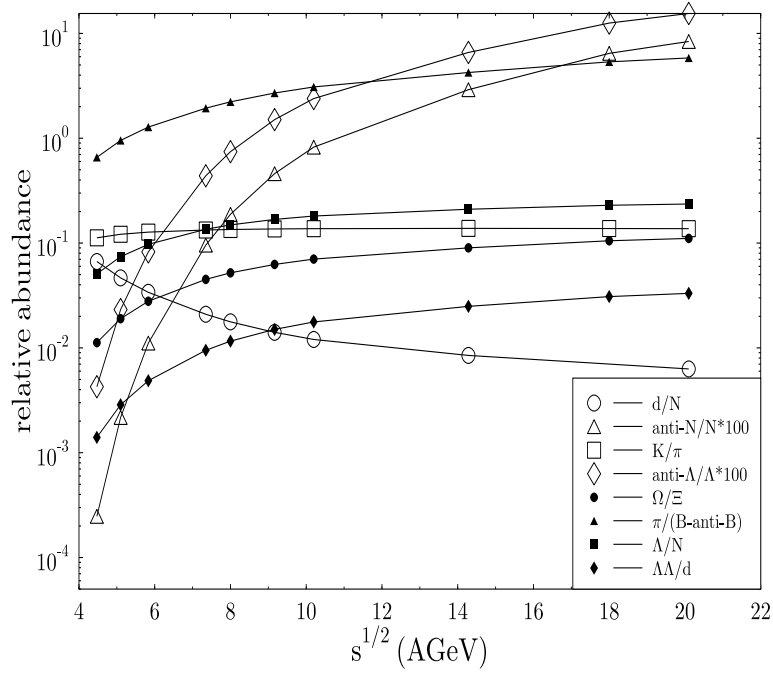,height=11cm,width=13cm}}}
\vspace{-1cm}
\caption{The excitation function of various particle ratios as calculated from
the S/A values obtained from the three-fluid
model. Feeding due to decays of resonances is taken into account.}
\label{fig_ratiosws}
\vspace{.5cm}
\end{figure}

The hadron ratios at AGS and SPS, cf.\ Fig.~\ref{fig_ratiosws}, are
quite close to the data discussed in the literature
\cite{PBMAGS,CSCG,sollfrank,letessier,bravina,marek}. For such a 
simple\footnote{We neglect e.g.\ that some hadron species might not reach
their chemical equilibrium abundance or decouple earlier (at higher
$\rho_B^{fo}$) than others. We also do not discuss here the rapidity
dependence of hadron ratios (cf.\ e.g.\ \cite{bass} for a calculation
within the microscopic transport model UrQMD) to study whether the
various contributions to ${\rm d}S/{\rm d}\eta$, as depicted in
Fig.~\ref{dsdy}, can be disentangled. This requires the one-fluid
solution in that part of the forward light-cone between the freeze-out
hypersurface and the hypersurface where local kinetic equilibrium
between the fluids is established. Such a calculation is out of the scope of
the present work.}
estimate of hadron production in nuclear collisions, deviations from
the experimental ratios by up to factors of two have to be expected.
Nevertheless, it is clear from Fig.~\ref{fig_ratiosws} that the simultaneous
measurement of various hadron ratios, like $\pi/\left(B-\overline{B}\right)$,
$d/N$ and, in particular, $\overline{B}/B$ (provided antibaryons also reach
chemical equilibrium) allows to determine the produced entropy in the
energy range between the AGS and the SPS. 
In contrast, the $K/\pi$-ratio is practically constant.

\section{Summary and outlook}

In the present paper we have calculated the amount of entropy produced
in the compression stage of relativistic heavy ion collisions using
three different macroscopic models. 

The three-fluid model is the only of these models which accounts for
(kinetic) non-equilibrium effects between the mutually interacting fluids.
Calculations within this model show that
the amount of entropy which is due to production of a third fluid (in addition
to the projectile and target fluids) of thermalized secondaries around
midrapidity can -- at higher bombarding energies -- exceed
the amount of entropy produced by nuclear shockwaves in the projectile
and target nuclei. The total specific entropy $S/A$ produced within
this model is consistent with the $S/A$ values extracted from data using 
relative particle yields from equilibrated hadron gases. We find, e.g.,
$S/A=11$ for AGS and $S/A=38$ for SPS energies.
The present calculations therefore support the picture
that most of the entropy production in relativistic heavy ion collisions
occurs in the early compression stage. Then a
(nearly) isentropic expansion follows.

The excitation function of the specific entropy $S/A(\sqrt{s})$
does not exhibit any threshold signatures of the phase transition to the QGP,
which is included in the EoS. Nevertheless, the phase
transition might still be visible in the excitation functions of other
observables, like e.g.\ directed in-plane flow.

The predicted particle yields exhibit a smooth 
excitation function, too. Simultaneous measurement of
various hadron ratios allows to
determine the amount of entropy produced in relativistic heavy ion
collisions.

Future work will also consider the rapidity and transverse momentum spectra of
various particles at freeze-out. A different EoS 
will be used during the expansion stage in order to account for
the chemical equilibration of additional particle species. The evolution
of the rapidity density of the entropy until freeze-out will be studied.

\section{Acknowledgements}

We are indebted to M.\ Ga\'zdzicki, M.I.\ Gorenstein, D.H.\ Rischke, D.\
R\"ohrich and R.\ Stock for helpful discussions. A.\ Dumitru
acknowledges a postdoctoral fellowship granted by the German Academic 
Exchange Service (DAAD).

%--------------------------------------------------------


\clearpage
\begin{thebibliography}{99}
\bibitem{entropy} W. Scheid, H. M\"uller, W. Greiner: Phys. Rev. Lett.
32 (1974) 741;\\
M.I. Sobel, P.J. Siemens, J.P. Bondorf, H.A. Bethe:
Nucl. Phys. A251 (1975) 502;\\
L.P. Csernai, J.I. Kapusta: Phys. Rep. 131 (1986) 223;\\
H. St\"ocker, W. Greiner:
Phys. Rep. 137 (1986) 277;\\
R.B. Clare, D. Strottman: Phys. Rep. 141 (1986) 177
\bibitem{harris} 
J. Harris et al.: Phys. Lett. B153 (1985) 377;\\
R. Stock: Phys. Rep. 135 (1986) 259
\bibitem{stoe} 
H. St\"ocker, W. Greiner, W. Scheid: Z. Phys. A286 (1978)
121;\\
H. St\"ocker: J. Phys. G10 (1984) L111;\\
D. Hahn, H. St\"ocker: Nucl. Phys. A476 (1988) 718
\bibitem{bertsch}
G. Bertsch, J. Cugnon: Phys. Rev. C24 (1981) 2514
\bibitem{csernai} 
I.M. Mishustin, F. Myhrer, P.J. Siemens: Phys. Lett. B95 (1980) 361;\\
L.P. Csernai, H.W. Barz: Z. Phys. A296 (1980) 173;\\
J. Kapusta: Phys. Rev. C24 (1981) 2545
\bibitem{PBMAGS} 
P. Braun-Munzinger, J. Stachel, J.P. Wessels, N. Xu:
Phys. Lett. B344 (1995) 43;
Phys. Lett. B365 (1996) 1
\bibitem{CSCG} 
C. Spieles, H. St\"ocker, C. Greiner: Eur. Phys. J. C2 (1998) 351;\\
A. Dumitru, C. Spieles, H. St\"ocker, C. Greiner: Phys. Rev. C56 (1997) 2202
\bibitem{sollfrank} 
J. Sollfrank: J.Phys. G23 (1997) 1903-1919 and references
therein
\bibitem{letessier} 
J. Letessier, A. Tounsi, U. Heinz, J. Sollfrank,
J. Rafelski: Phys. Rev. Lett. 70 (1993) 3530;\\
E. Suhonen, J. Cleymans, K. Redlich, H. Satz: hep-ph/9310345;\\
J. Cleymans, D. Elliott, H. Satz, R.L. Thews: Z. Phys. C74 (1997) 319;\\
G.D. Yen, M.I. Gorenstein, W. Greiner, S.N. Yang: Phys. Rev. C56 (1997) 2210
\bibitem{bravina}
L.V. Bravina et al.: nucl-th/9804008
\bibitem{sollfrank2} 
J. Sollfrank, P. Huovinen, M. Kataja, P.V. Ruuskanen, M. Prakash,
R. Venugopalan: Phys. Rev. C55 (1997) 392;\\
C.M. Hung, E. Shuryak: Phys. Rev. C57 (1998) 1891
\bibitem{marek} 
M. Gazdzicki, D. R\"ohrich:  Z. Phys. C65 (1995) 215;\\
J. G\"unther: PhD Thesis, Univ. Frankfurt, 1997
\bibitem{sigom}
M.I. Gorenstein, D.H. Rischke, H. St\"ocker, W. Greiner: J. Phys. G19
(1993) L69
\bibitem{Bagmodel} 
A. Chodos, R.L. Jaffe, K. Johnson, C.B. Thorn,
V. Weisskopf: Phys. Rev. D9 (1974) 3471
\bibitem{Kap79} 
S.A. Chin: Phys. Lett. B78 (1978) 552;\\
J. Kapusta: Nucl. Phys. B148 (1979) 461;\\
E.V. Shuryak: Phys. Rep. 61 (1980) 71;\\
J. Cleymans, R.V. Gavai, E. Suhonen: Phys. Rep. 130 (1986) 217
\bibitem{RiShock} 
D.H. Rischke, Y. P\"urs\"un, J.A. Maruhn:
Nucl. Phys. A595 (1995) 383
\bibitem{danielewicz}
P. Danielewicz: Phys. Lett. B146 (1984) 168
\bibitem{Landau} 
L.D. Landau, E.M. Lifshitz: ``Fluid Mechanics'',
Pergamon Press, New York, 1959
\bibitem{Bugaev} 
K.A. Bugaev, M.I. Gorenstein, V.I. Zhdanov: Z. Phys. C39 (1988) 365;\\
K.A. Bugaev, M.I. Gorenstein, B. K\"ampfer, V.I. Zhdanov:
Phys. Rev. D40 (1989) 2903;\\
K.A. Bugaev, M.I. Gorenstein, D.H. Rischke: Phys. Lett. B255 (1991) 18;\\
D.H. Rischke, Y. P\"urs\"un, J.A. Maruhn, H. St\"ocker, W. Greiner:
Heavy Ion Phys. 1 (1995) 309
\bibitem{Hofmann} 
J. Hofmann, H. St\"ocker, U. Heinz, W. Scheid, W. Greiner:
Phys. Rev. Lett. 36 (1976) 88
\bibitem{Taub} 
A.M. Taub: Phys. Rev. 74 (1948) 328;\\
H.G. Baumgardt et al.: Z. Phys. A237 (1975) 359;\\
J.R. Nix: Prog. Part. Nucl. Phys. 2 (1979) 237;\\
H. St\"ocker, M. Gyulassy, J. Boguta: Phys. Lett. B103 (1981) 269
\bibitem{TwoFl} 
A.A. Amsden, A.S. Goldhaber, F.H. Harlow,
J.R. Nix: Phys. Rev. C17 (1978) 2080;\\  
L.P. Csernai, I. Lovas, J.A. Maruhn, A. Rosenhauer, J. Zim\'anyi, 
W. Greiner: Phys. Rev. C26 (1982) 149;\\
H.W. Barz, B. K\"ampfer, L.P. Csernai, B. Lukacs:
     Nucl. Phys. A465 (1987) 743;\\
H.W. Barz, B. K\"ampfer: Phys. Lett. B206 (1988) 399
\bibitem{pp2fluid} 
I.N.~Mishustin, V.N.~Russkikh, L.M.~Satarov:   
Sov. J. Nucl. Phys. 48 (1988) 454; Nucl. Phys. A494 (1989) 595;\\
L.M.~Satarov: Sov. J. Nucl. Phys. 52 (1990) 264;\\
I.N. Mishustin, L.M. Satarov,  V.N. Russkikh: in ``Relativistic Heavy Ion
Physics'' (eds. D. Strottman and L.P. Csernai), vol. 1,
World Scientific (Singapore), 1991, p.179 and
Sov. J. Nucl. Phys. 54 (1991) 459
\bibitem{BrachiNPA} 
J. Brachmann, A. Dumitru, J.A. Maruhn, H. St\"ocker,
W. Greiner, D.H. Rischke: Nucl. Phys. A619 (1997) 391
\bibitem{Trento} 
A. Dumitru, J. Brachmann, M. Bleicher, J.A. Maruhn,
H. St\"ocker, W. Greiner: Heavy Ion Phys. 5 (1997) 357
\bibitem{RiFri} 
D.H. Rischke, B.L. Friman, B.M. Waldhauser, H. St\"ocker,
W. Greiner: Phys. Rev. D41 (1990) 111
\bibitem{PRCZPA} 
A.~Dumitru, U.~Katscher, J.A.~Maruhn, H.~St\"ocker,
W.~Greiner, D.H.~Rischke: Phys. Rev. C51 (1995) 2166
\bibitem{Blobel} 
V. Blobel et al.: Nucl. Phys. B69 (1974) 454
\bibitem{anishetty}
R. Anishetty, P. Koehler, L. McLerran: Phys. Rev. D22 (1980) 2793;\\
J.D. Bjorken: Phys. Rev. D27 (1983) 140
\bibitem{antinucci} 
M. Antinucci, A. Bertin, P. Capiluppi, M. d'Agostino-Bruno et al.: 
Lett. Nuov. Cim. 6 (1973) 121
\bibitem{bass}
S.A. Bass et al.: nucl-th/9711032
\end{thebibliography}
\end{document}